\documentclass[referee,traditabstract]{aa} 
\usepackage{graphicx}
\usepackage{natbib}
\bibpunct{(}{)}{;}{a}{}{,} 
\usepackage{txfonts}
\newcommand{\mr}[1]{\mathrm{#1}}

\def\<{\left\langle} \def\>{\right\rangle} \def\({\left(} \def\){\right)}
\def\[{\left[}
\def\]{\right]}
\def\Gi{G_\infty}
\def\et{\textit{et al. }}

\def\RB{Rayleigh-B\'{e}nard }
\def\TC{Taylor-Couette }

\def\N0{N^{\omega}_0 }

\def\Rom{R_{\Omega} }

\def\vK{von K\'{a}rm\'{a}n }
\def\PvK{Prandtl--von K\'{a}rm\'{a}n }

\def\EQ{\begin{equation}}
\def\EN{\end{equation}}
\def\EQA{\begin{eqnarray}}
\def\ENA{\end{eqnarray}}

\begin{document}
   \title{Angular momentum transport and turbulence in laboratory models of Keplerian flows}

   \author{M.~S.~Paoletti
          \inst{1}
          \and
          Dennis P.M.~van Gils\inst{2}
          \and
          B.~Dubrulle\inst{3}
          \and
          Chao Sun\inst{2}
          \and
          Detlef Lohse\inst{2}
          \and
          D.~P.~Lathrop\inst{4}
          }

   \institute{Department of Physics and Center for Nonlinear Dynamics,\\
             University of Texas, Austin, TX 78712
        \and
             Physics of Fluids Group, Faculty of Science and Technology, Mesa+ Institute and Burgers Center for Fluid Dynamics, University of Twente, 7500AE Enschede, The Netherlands
        \and
            CNRS URA 2464 GIT/SPEC/IRAMIS/DSM, CEA Saclay, F-91191 Gif-sur-Yvette, France
        \and
            Departments of Physics and Geology, Institute for Research in Electronics and Applied Physics, Institute for Physical Science and Technology, University of Maryland, College Park, MD 20742
             \email{lathrop@umd.edu}
             }

   \date{Received ??, 2011; accepted ??, ??}

\abstract{We present angular momentum transport (torque) measurements in two recent experimental studies of the turbulent flow between independently rotating cylinders.  In addition to these studies, we reanalyze prior torque measurements to expand the range of control parameters for the experimental Taylor-Couette flows.  We find that the torque may be described as a product of functions that depend only on the Reynolds number, which describes the turbulent driving intensity, and the rotation number, which characterizes the effects of global rotation.  For a given Reynolds number, the global angular momentum transport for Keplerian-like flow profiles is approximately 14\% of the maximum achievable transport rate.  We estimate that this level of transport would produce an accretion rate of $\dot{M}/\dot{M_0} \sim 10^{-3}$ in astrophysical disks.  We argue that this level of transport from hydrodynamics alone could be significant.}

   \titlerunning{Angular momentum transport and turbulence in Keplerian flows}
   \authorrunning{Paoletti \et}
   \maketitle

%

\section{Introduction}
Astrophysical disks are ubiquitous in the universe, orbiting around a wide panel of massive objects, from black holes to stars. An essential ingredient of their dynamics is the radial transport of angular momentum that governs the rate of material falling onto the central object. Realistic models  of this process are impeded by the complexity of the disk dynamics, including a wide variety of processes such as general relativistic effects, self-gravitation, radiation, dynamics of plasma magnetic fields and turbulence.  In particular, the contribution from hydrodynamics to the angular momentum transport in such astrophysical flows is not yet understood.

A natural first-step in the construction of a robust model is to consider the minimal list of ingredients that can capture and reproduce the observed disk properties. Such a model has been proposed several decades ago by \citet{shakura73} and \citet{pringle81}, and was given the name $\alpha$ disks. In that model, the radial transport of angular momentum is parameterized by a turbulent viscosity coefficient, that needs to be prescribed, or measured.  In the past, there have been a few attempts to measure this coefficient in numerical simulations \citep{dubrulle92}.

An alternative promising option, as recognized by \citet{zeldovich81} and \citet{richard99}, is offered by focusing instead on laboratory experiments. Indeed, it may be shown \citep{hersant05}, that under simple, but well founded approximations, the equations governing an $\alpha$ disk are similar to the equation of motion of an incompressible rotating shear flow, with penetrable boundary conditions and cylindrical geometry. This kind of flow can be achieved in the Taylor-Couette flow (see Fig.\ \ref{TC_schematic}), a fluid layer sheared between two independently rotating, coaxial cylinders.

This remark motivated several experimental studies of the transport properties in the Taylor-Couette flow, with contradicting conclusions.
Depending upon the community, the strength of the turbulent transport is quantified in various ways.
The most basic quantification \citep{wendt33} is the one through the torque $T$ which is necessary
to keep the inner cylinder rotating at a given velocity. The dimensionless version thereof is
$G = T / (L_{mid} \rho \nu^2 ) $, where $\rho$ is the fluid density, $\nu$ its kinematic viscosity
and $L_{mid} $ the effective length of the cylinders \citep{paoletti11}. \citep{eckhardt07a} and \citep{vangils11} used
$Nu_\omega = G/ G_{\mr{lam}}$, where $G_{\mr{lam}}$ is the laminar torque, to quantify the angular velocity flux from the inner to the outer cylinder, in order to highlight the analogy between \TC and \RB flow \citep{dubrulle02,eckhardt07a,eckhardt07b}.
The engineering community prefers the drag-coefficient $c_f = G/ Re^2$, where $Re = {2\over 1+\eta} |\eta Re_2 - Re_1|$ and
$\eta = a/b$ is the radius ratio and $Re_{1/2} $ the Reynolds number of the inner/outer cylinder.
In the astrophysics community, the same information is often expressed in terms of
the $\beta$ coefficient, defined by $\beta = 2 G \eta^2 / (\pi (1-\eta)^4 Re^2 ) = 2 \eta^2 c_f /  (\pi (1-\eta)^4  ) $, which can be interpreted as a dimensionless turbulent viscosity.

By reexamining previous data obtained by Wendt \citep{wendt33} and Taylor \citep{taylor36}, \citet{richard99}
showed that the $\beta$-parameter is
on the order of $\beta_{RZ}\sim 10^{-5}$ for Keplerian disks.
A more complete analysis, including results of new experiments on a classical Taylor-Couette flow from \citet{richardthesis}, led to the same conclusion \citep{dubrulle05}. In contrast, an original experiment, built in Princeton \citep{ji06} so as to minimize end effects, measured a $\beta$ coefficient through radial Reynolds stresses and obtained a value smaller by more than an order of magnitude $\beta_P\sim 7 \times 10^{-7}$.  Furthermore, the Princeton experiments could not distinguish the angular momentum transport in Keplerian flows from their measurements in solid-body rotation, where the cylinders rotate together thereby resulting in zero angular momentum transport.

In the present contribution, we report precise angular momentum transport (torque) measurements in two independent experiments performed at very large Reynolds number and compute the resulting $\beta$ coefficient for different rotation profiles.  To further study the scaling of angular momentum transport in rotating shear flows, we compare our measurements to those of \citet{wendt33}, \citet{taylor36} and \citet{richardthesis,dubrulle05}.  We find a universal scaling of the torque for various rotation profiles that captures the effects of the various geometries used in the experiments summarized here.

\section{Apparatus and experimental details}

\subsection{Apparatus: generalities}
\label{Apparatus}

\begin{figure}[tb]
\begin{center}
 \includegraphics[width=0.5\columnwidth]{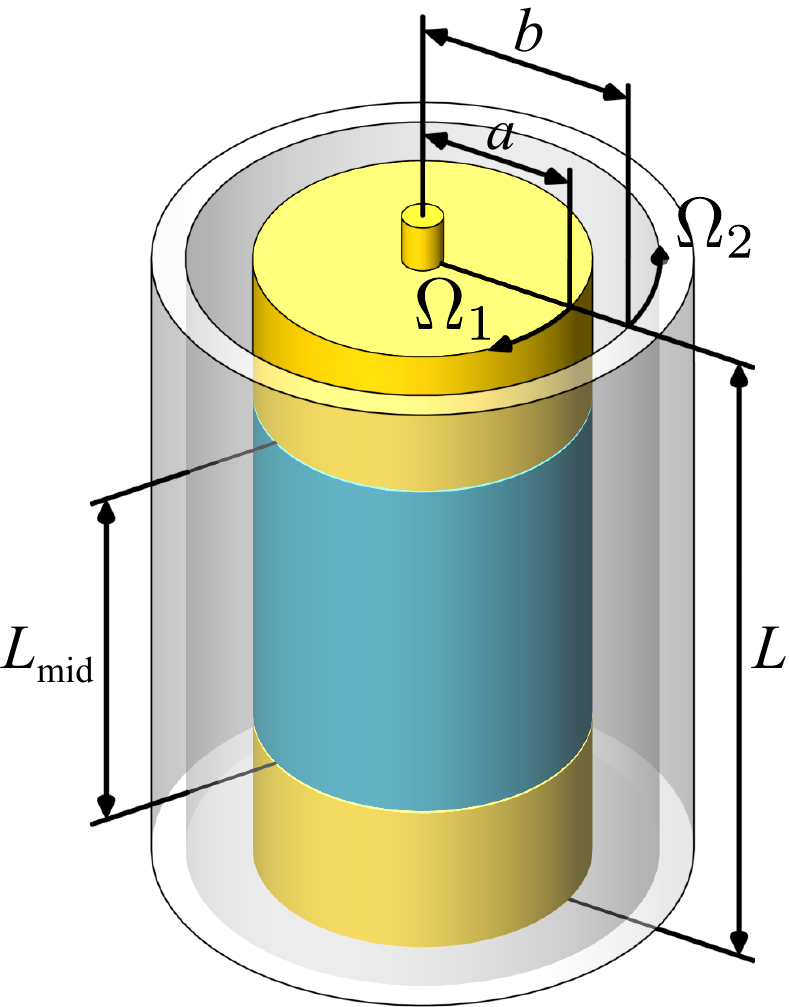}%
\caption{Taylor-Couette flow may be used to study angular momentum transport in rotating shear flows.  The fluid is sheared between two independently rotating cylinders of radii $a < b$ that rotate at angular velocities $\Omega_1$ and $\Omega_2$ for the inner and outer cylinders, respectively.  Although the total fluid height is given by $L$, some experiments only measure the contribution to the torque over a central section of length $L_{\mr{mid}}$ to minimize end effects, as in  \citep{lathrop92a,lathrop92b,lewis99,vandenBerg03,vandenBerg05,vandenBerg07,paoletti11,vangils11}.}
\label{TC_schematic}
\end{center}
\end{figure}

The experiments presented here all examine the transport of angular momentum (torque) in classical Taylor-Couette flow.  Figure \ref{TC_schematic} shows a schematic representation of our experiments.  The fluid is contained between concentric cylinders of radii $a < b$.  The inner (outer) cylinder rotates at an angular velocity of $\Omega_1 \( \Omega_2\)$.  The overall height of the fluid contained between the axial boundaries is given by the length $L$.  The geometry of a particular Taylor-Couette flow apparatus is often defined in dimensionless form by the radius ratio $\eta \equiv a/b$ and the aspect ratio $\Gamma \equiv L/(b-a)$.

The ideal Couette flow is infinite in axial extent.  As such, several methods have been employed to handle the finite-size effects that are present in any laboratory experiment.  The most common case has the axial boundaries rotate with one of the two cylinders.  In other experiments, the axial boundaries are divided at the mid-gap and have the inner (outer) portion rotate with the inner (outer) cylinder.  This split-ring approach has been further developed to allow for pairs of rings that can independently rotate with respect to the cylinders.  Allowing the axial boundaries to independently rotate aids in the suppression of finite-size effects as detailed in  \citep{ji06,schartman09,burin10}.

\begin{table*}
\caption{A summary of the parameters used in the experiments discussed here.  The geometry of the apparatus is given by the outer cylinder radius $b$, the radius ratio $\eta = a/b$ and the aspect ratio $\Gamma = L/(b-a)$.  The range of turbulence intensities are governed by the inner and outer cylinder Reynolds numbers $Re_1$ and $Re_2$.  The effects of rotation are parameterized by the rotation number $\Rom$ introduced by \citet{dubrulle05}.}
\centering
\begin{tabular*}{\textwidth}{@{\extracolsep{\fill}}cccccccc}
\hline\hline
Experiment & $b$~(cm) & $\eta$ & $\Gamma$ & max($Re_1$) & max($Re_2$) & min($\Rom$) & max($\Rom$)\\
\hline
\citet{wendt33} & 14.7 & 0.68 & 10.64 & $8.47 \times 10^4$ & $1.24 \times 10^5$ & -0.71 & 0.47\\
\citet{wendt33} & 14.7 & 0.85 & 22.93 & $4.95 \times 10^4$ & $5.83 \times 10^4$ & -0.80 & 0.18\\
\citet{wendt33} & 14.7 & 0.935 & 52.63 & $2.35 \times 10^4$ & $2.52 \times 10^4$ & -0.42 & 0.11\\
\citet{taylor36} & 8.11 & 0.79--0.973 & 50--384 & 0 & $1.08 \times 10^5$ & 0.03 & 0.27\\
\citet{richardthesis} & 5.00 & 0.72 & 27.18 & $9.50 \times 10^4$ & $1.30 \times 10^5$ & -1.43 & -0.94\\
\citet{richardthesis} & 5.00 & 0.72 & 27.18 & $9.50 \times 10^4$ & $1.30 \times 10^5$ & 0.38 & 0.58\\
\citet{ji06} & 20.3 & 0.35 & 2.1 & $1.20 \times 10^6$ & $4.50 \times 10^5$ & -1.67 & -1.08\\
Maryland & 22.09 & 0.725 & 11.47 & $3.70 \times 10^6$ & $1.50 \times 10^6$ & -2.04 & 1.46\\
Twente & 27.9 & 0.716 &  11.68 & $2.00 \times 10^6$ & $1.40 \times 10^6$ & -0.74 & 0.17\\
\hline
\end{tabular*}
\label{Table_params}
\end{table*}

The data that we analyze and present stems from several different experiments, which are summarized in Table \ref{Table_params}.  The early experiments by \citet{wendt33} measured the torque required to drive the inner cylinder for three values of $\eta$.  The system had a free surface at the top, although we quote an effective aspect ratio defined using the axial length of the apparatus.  The data from \citet{taylor36} determined the scaling of the torque for several values of $\eta$ with the inner cylinder stationary ($\Omega_1 = 0$).  Richard's experiments \citep{richardthesis, dubrulle05} did not directly measure the torque.  Instead, critical Reynolds numbers for various transitions were observed and the analysis presented by \citet{dubrulle05} was used to estimate the corresponding torque.  We do not present data from the experiments of \citet{ji06}, but rather use their reported measurements as a basis for comparison, as detailed in Section \ref{Discussion}.  The two most recent experiments by \citet{paoletti11} and \citet{vangils11} are further described in the forthcoming sections.

\subsection{Apparatus: the Maryland experiment}
 The experiments presented by \citet{paoletti11}, henceforth referred to as the Maryland experiment, are conducted in the apparatus constructed by \citet{lathrop92a,lathrop92b}, which was modified to allow the outer cylinder to rotate independently.  The outer cylinder is an anodized aluminum cylinder with the same physical dimensions as in \citet{lathrop92a,lathrop92b,lewis99}, specifically $b = 22.085$~cm and a working fluid height $L = 69.50$~cm.  The inner cylinder is stainless steel with a radius $a = 16.000$~cm yielding a radius ratio $\eta = 0.7245$ and an aspect ratio $\Gamma = 11.47$.  The inner cylinder is rotated up to $\Omega_1/2\pi = 20$~s$^{-1}$ while the outer cylinder may be rotated in either direction up to $\left|\Omega_2/2\pi\right| = 10$~s$^{-1}$.  Both angular velocities are measured by shaft encoders and controlled to within 0.2\% of the set value.

The axial boundaries in the Maryland experiment rotate with the outer cylinder.  To reduce end effects in the torque measurements the inner cylinder is divided axially into three sections of length 15.69, 40.64 and 15.69~cm (see schematics in \citet{lathrop92a,lathrop92b}).  Only the central section of length $L_{\mr{mid}} = 40.64$~cm (see Fig.\ \ref{TC_schematic}) senses the torque of the fluid as described in  \citet{lathrop92a,lathrop92b}.  The regions $2.58$ gap widths from each of the axial boundaries, where secondary circulation setup by finite boundaries is strongest, are avoided in the torque measurements.

The desired accuracy of our measurements requires that the temperature of the water be precisely controlled.  In contrast to prior experiments \citep{lathrop92a,lathrop92b,lewis99,vandenBerg03,vandenBerg05,vandenBerg07,vangils11} where the system was cooled at the axial boundaries, we control the temperature of the water by heating and cooling the outer cylinder.  This allows the working fluid to be temperature-controlled along the entire axial length of the experiment, yielding a 6.5 fold increase in the temperature-controlled surface area.  This is particularly important for Rayleigh-stable flows, where mixing is greatly reduced.  In our experiments the temperature is $50 \pm 0.02 ^{\circ}$C yielding a kinematic fluid viscosity of $\nu = 5.5 \times 10^{-3}$~cm$^2$/s, except for $Re > 2 \times 10^{6}$ where $T = 90$~$^{\circ}$C and $\nu = 3.2 \times 10^{-3}$~cm$^2$/s.

\subsection{Apparatus: the Twente experiment}
The Twente turbulent \TC (called T$^3$C) facility, here referred to as the Twente experiment, is described in great detail in \citet{vangils11b}. In short, the working fluid height is $L = 92.7$~cm, has an inner radius of $a = 20.0$~cm and an outer radius of $b = 27.94$~cm.  The maximum inner and outer angular velocities are $\Omega_1/2\pi = 20$~Hz and $\left|\Omega_2/2\pi \right| = 10$~Hz, respectively.
The system is fully temperature controlled through cooling of the upper and lower end plates, which co-rotate with the outer cylinder. The system is operated with water at a temperature of $20$~$^{\circ}$C, resulting in a
kinematic viscosity of $\nu = 1.04 \times 10^{-2}$~cm$^2$/s.
The torque is measured over the middle part of the inner cylinder of height $L_{\mr{mid}} = 53.6$~cm to minimize the influence of the end-plates (similar to \citep{lathrop92a,lathrop92b}). The torque is measured by a load cell imbedded inside the middle section of the inner cylinder in both the Twente and Maryland experiments, in contrast to measuring the torque on the drive shaft that would be affected by mechanical seals in the end plates or by velocimetry measurements, as in the Princeton experiments \citep{ji06,schartman09,burin10,schartman11}.

\subsection{Control parameters}
The analysis of our results is simplified by a proper choice of the control parameters. As in \citet{dubrulle05}, we use  the following three dimensionless control parameters, which fully
describe the geometry and specify the state of the dynamical \TC system:
\EQA
Re&=&
\frac{2}{1+\eta}\vert \eta Re_2 - Re_1\vert, \label{re} \\
R_\Omega&=&
(1-\eta)\frac{Re_1 + Re_2}{\eta Re_2 - Re_1}, \label{re-omega}\\
R_{\cal C}&=&\frac{1-\eta}{\eta^{1/2}},
\label{controlTC}
\ENA
where $Re_1=a\Omega_1d/\nu$ and $Re_2=b \Omega_2d/\nu$ are the Reynolds number of the inner and outer cylinder and $d = (b-a)$ is the gap width.

The above control parameters have been introduced so that their definitions apply to rotating shear flows in general and not strictly only
to the Taylor-Couette geometry. It is very easy in this formulation to relate the Taylor-Couette flow to the shearing sheet (plane Couette flow with rotation), by considering the limit $R_{\cal C}\rightarrow 0$. The linear stability properties of the fluid can also be neatly recast using these control parameters.
In the inviscid limit ($Re\rightarrow\infty$), and for axisymmetric disturbances, the linear stability properties of the flow are governed by the Rayleigh criterion, namely that  the fluid is stable if the Rayleigh discriminant is everywhere positive:
\begin{equation}
\frac{\Omega}{r} \partial_r L(r)>0,
\label{Rayleigh}
\end{equation}
where $L(r)=r^2\Omega(r)$ is the specific angular momentum. Applying this criterion to the laminar profile leads to
\begin{equation}
(R_\Omega+1)(R_\Omega+1- ab /r^2)>0.
\label{newrayleigh}
\end{equation}

\noindent
Since $ab/r^2$ varies between $1/\eta$ and $\eta$, one obtains that in the inviscid limit, the flow is unstable against infinitesimal axisymmetric disturbances when ${R_{\Omega}^{\infty}}^{-}<R_{\Omega}<{R_{\Omega}^{\infty}}^{+}$, where ${R_{\Omega}^{\infty}}^{-}=-1$, respectively ${R_{\Omega}^{\infty}}^{+}=1/\eta-1$, are the marginal stability thresholds in the inviscid limit (superscript $\infty$) in the cyclonic case ($R_\Omega>0$, subscript $+$), and anticyclonic case ($R_\Omega<0$, subscript $-$). This means that, in the anticyclonic case, any flow such that $R_\Omega<-1$ is linearly stable against axisymmetric disturbances. However, it may be non-linearly stable or unstable against non-axisymmetric disturbances, as found in the experiments described here.\

In the present paper, we determine the intensity of the momentum transport by measuring the torque required to drive the inner cylinder for a given set of parameters $\(Re, \Rom\)$. Measurements described here suggest that at large enough Reynolds number, the nondimensional torque $G = T/ (\rho \nu^2 L_{\mr{mid}})$, where $T$ is the torque, $\rho$ the fluid density, $\nu$ the kinematic viscosity and $L_{\mr{mid}}$ is defined in Fig.\ \ref{TC_schematic}, varies approximately as $G\sim Re^{\alpha}$ with $1.75 < \alpha(Re) < 1.85$, even for Rayleigh-stable flows. While this scaling exceeds the laminar case $(G \sim Re)$, we note that these exponents are less than a simple dimensional argument would give: Such an argument would say that in a fully turbulent regime the transport properties could not depend on the molecular viscosity, which would imply
 that the quantity $c_f \sim G/Re^2$ is Reynolds number independent. The deviations from $G\sim Re^2$
mean that the viscosity is always relevant for the transport, due to its dominance in the boundary layers.
Only when the boundary layers are sufficiently disturbed by introducing wall
roughness does one obtain $G\sim Re^2$,
as we have shown in our previous work \citep{vandenBerg03}.

\begin{table*}
\caption{A summary of the Taylor-Couette control parameters related to some astrophysical disks, as computed by \citet{hersant05}. Here, $b$ is the disk's outer radius, and $\dot M$ the accretion rate.}
\centering
\begin{tabular*}{\textwidth}{@{\extracolsep{\fill}}lccccccc}
\hline\hline
Central object & $b$~(a.u.) & $\eta$ & $\Gamma$ & max($Re_1$) & max($Re_2$) & $\Rom$ & $\dot{M}/\dot{M_0}$\\
\hline
TTauri  & 1000 & 0 & 0.94 & $10^{12}$ & $10^{14}$ & -1.33 & $10^{-5}-10^{-2}$\\
FuOri & 5 & 0 & 0.94 & $10^{13}$ & $10^{15}$ & -1.33 & 0.1-1\\
\hline
\end{tabular*}
\label{Table_astro}
\end{table*}

\subsection{Astrophysically relevant quantity}
\label{Astrophysically relevant quantity}
In the astrophysical context, one often considers asymptotic angular velocity profiles of the form $\Omega(r)\sim r^{-q}$ where $q$ characterizes the flow. In that case $q=-\partial\ln\Omega/\partial\ln r=-2/R_{\Omega}$, which relates astrophysical profiles in the control parameters space of Taylor-Couette flows. For Keplerian flow, $q=3/2$ resulting in $R_\Omega=-4/3$, which is an example of a linearly stable, anticyclonic flow.  While Keplerian profiles cannot be precisely achieved by Taylor-Couette flows, states with $\Rom < -1$ are all linearly stable, anticyclonic flows that may be used to approximate astrophysical settings.  We will refer to cases with $\Rom < -1$ as quasi-Keplerian, as in Ref.\ \citep{ji06}.

Astrophysical disks are very wide with $a\ll b$. Therefore, their geometrical global value of $R_{\cal C}\to\infty$. In a {\sl local} description of disks, one often considers the shearing sheet approximation, in which $\eta\to 1$, so  that $R_{\cal C}\to 0$, as in plane-Couette flow. Here, we focus on global properties and consider only the limit $\eta \rightarrow 0$ applicable to disks in the discussion.  As a result of their astronomical size, the Reynolds numbers associated with accretion disks are very large. To give a few specific examples, parameters for typical TTauri and FuOri stars have been computed by \citet{hersant05} and are given in Table \ref{Table_astro}.

In disks,  angular momentum transport intensity is quantified through the accretion rate $\dot{M}$.
In fact, as was shown in \citep{dubrulle05,hersant05}, this rate is directly related to the turbulent viscosity and the dimensionless torque.  This concept may be understood in the following way: pressure gradients substantially balance centrifugal forces that develop in (incompressible) laboratory flows.  Keplerian disks, however, balance the centrifugal force with gravitational attraction.  Thus when angular momentum is transported radially outward, accretion disk matter responds by falling radially inward at the rate $\dot{M}$.  The flux of angular momentum, which is proportional to $G$, is thereby related to the accretion rate $\dot{M}$ by \citep{dubrulle05,hersant05},
\begin{equation}
\frac{G}{Re^2}=\frac{\pi}{2R_{\cal C}^4}\beta=\frac{\dot M}{\dot M_0},
 \label{considerons}
 \end{equation}
 where $\beta$ is a turbulent viscosity parameter such that $\nu_t=\beta S r^2$, with $S$ a velocity shear, and
 $\dot M_0$ is an effective accretion rate given by the mean surface density $\bar{\Sigma}$, the disk inner radius $a$, mean radius $\bar{r} = (a+b)/2$,the mean angular velocity $\bar{\Omega}$ and the mean disk height $\bar H$ through
 \begin{equation}
 \dot{M_{0}}=\bar{\Sigma} a \bar{r} \bar{\Omega}\left(\frac{\bar{H}}{\bar{r}}\right)^4.
 \label{utileMast}
  \end{equation}
 As such, measurements of $G/Re^2$ in laboratory experiments may be used to estimate the hydrodynamical contribution to angular momentum transport in accretion disks.

\begin{figure}[tb]
\begin{center}
\includegraphics[width=0.5\columnwidth]{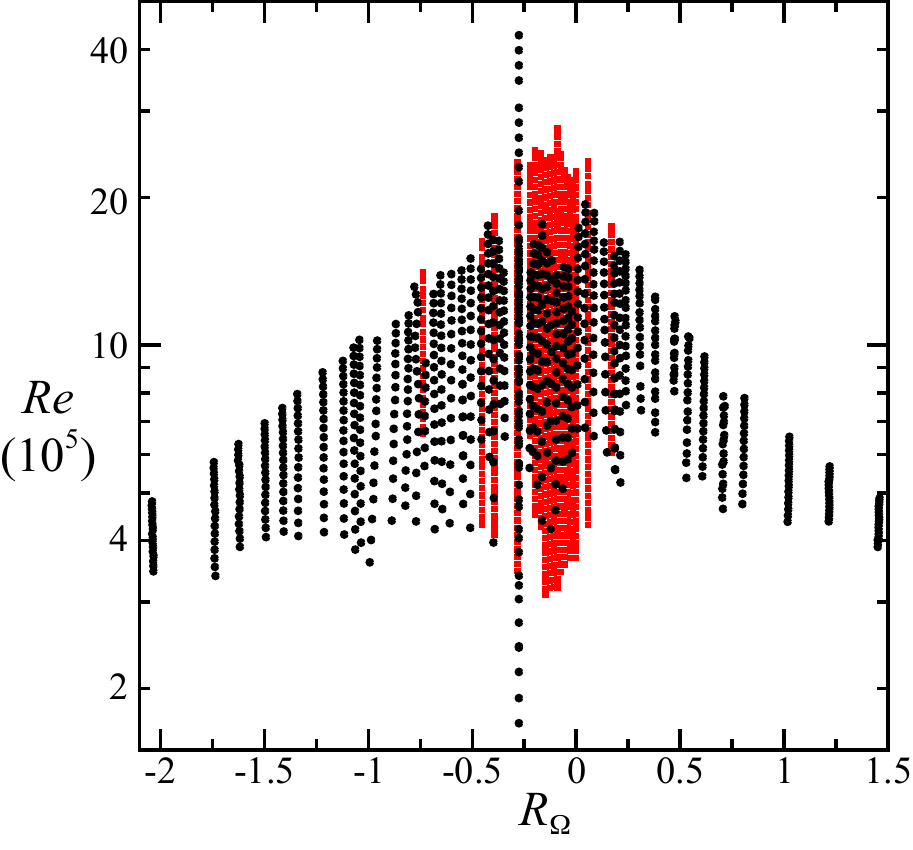}%
\caption{Our experiments span a large range of the Reynolds number $Re$, which controls the turbulence intensity, and the rotation number $\Rom$, introduced by Dubrulle \et \citep{dubrulle05}, which compares shear to overall rotation.  The rotation number can be used to distinguish between Rayleigh-stable flows ($\Rom \le -1$ or $\Rom \le (1-\eta)/\eta$) and those that are linearly unstable $(-1 < \Rom < (1-\eta)/\eta)$.  The data are taken from separate experiments by \citet{vangils11} (red squares) and \citep{paoletti11} (black circles).}
\label{Re_vs_Rom2}
\end{center}
\end{figure}

\section{Results}
We study the scaling of the dimensionless torque $G$ over a range of Reynolds numbers for various constant rotation numbers (see Fig.\ \ref{Re_vs_Rom2}), spanning a wide variety of flow states, including linearly stable cyclonic and anticyclonic flows as well as linearly unstable states.  Corresponding torque measurements from the Twente experiment are shown as red squares in Fig.\ \ref{G_vs_Re} while the data from Maryland are represented by black circles.

\subsection{Existence of turbulence}

\begin{figure}[tb]
\begin{center}
\includegraphics[width=0.5\columnwidth]{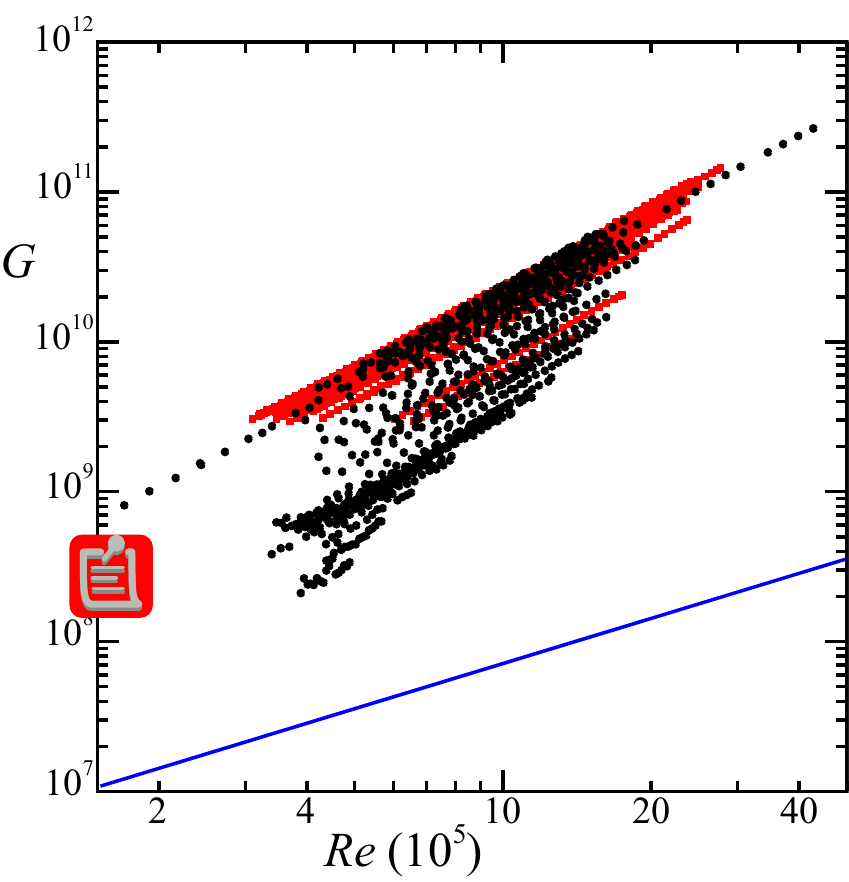}%
\caption{The scaling of the dimensionless torque $G$ with the Reynolds number $Re$ is independent of the rotation number $\Rom$, however the amplitude varies as in Refs.\ \citep{paoletti11,vangils11}.  The measurements by \citet{paoletti11} are in black circles while the data from \citet{vangils11} are shown by red squares. The thick blue line indicates the theoretical value for purely laminar flow, where angular momentum is transported only by molecular viscosity.  }
\label{G_vs_Re}
\end{center}
\end{figure}

In laminar flow, angular momentum is transported only by molecular viscosity, so that the dimensionless torque $G$ is proportional to $Re$, with the prefactor depending upon $R_\Omega$. In Fig.\ \ref{G_vs_Re}, we show the expected laminar angular momentum transport (blue line) for the range of $Re$ explored in this study for the case $\Omega_2 = 0$. As can be seen all the torque measurements considered differ from the theoretical expectation for laminar flow in two ways: (\textit{i}) they increase faster than $Re$ and  (\textit{ii}) they are much higher than the theoretical laminar value. This is an indication that the fluid is turbulent, with enhanced angular momentum transport in all the cases reported here. Note that these cases encompass several situations with $R_\Omega<-1$ (linearly stable anticyclonic flow) and $R_\Omega>0.4$ (linearly stable cyclonic flow) in addition to many unstable states.

In particular, (approximate) Keplerian flows at $R_\Omega=-4/3$ are turbulent at $Re\sim 10^5$ in our experiments. This is in agreement with direct, visual observations of \citet{richardthesis}, performed at somewhat lower Reynolds number. We argue that this indicates that quasi-Keplerian flows can efficiently transport angular momentum in spite of their linear stability.  However, the nature of this nonlinear instability remains unclear.  We suggest that systematically perturbing quasi-Keplerian flow states at lower values of $Re$ while measuring the torque, as in the experiments by \citet{taylor36}, could aid in the description of this likely nonlinear instability.

\subsection{Enhancement of momentum transport}
We now try to  quantify the enhancement of angular momentum transport with respect to its laminar value, as a function of both $Re$ and $R_\Omega$.  As in
 \citet{paoletti11,vangils11}, the torque at a given $Re$ may be either increased or reduced depending upon $\Rom$; the Reynolds number alone is insufficient to describe the transport.

\subsubsection{Variation with $R_\Omega$}
\citet{paoletti11} observed that variation of the
torque at a given $Re$ shows a pronounced maximum as a function of $R_\Omega$.  An analogous dependence of the amplitude of an effective power-law scaling was also observed by van Gils \et \citep{vangils11}.  To determine the effects of global rotation, we normalize $G$ for each $Re$
by $G(Re, R_\Omega = 0)$ \citep{dubrulle05}, which we denote as $G_0$.  The dependence of $G/G_0$
on $R_\Omega$ is shown in Fig.\ \ref{Gcomp_vs_Re_UMD_Twente} with the data from Maryland shown as black circles and Twente in red squares.

\begin{figure}
\begin{center}
\includegraphics[width=0.5\columnwidth]{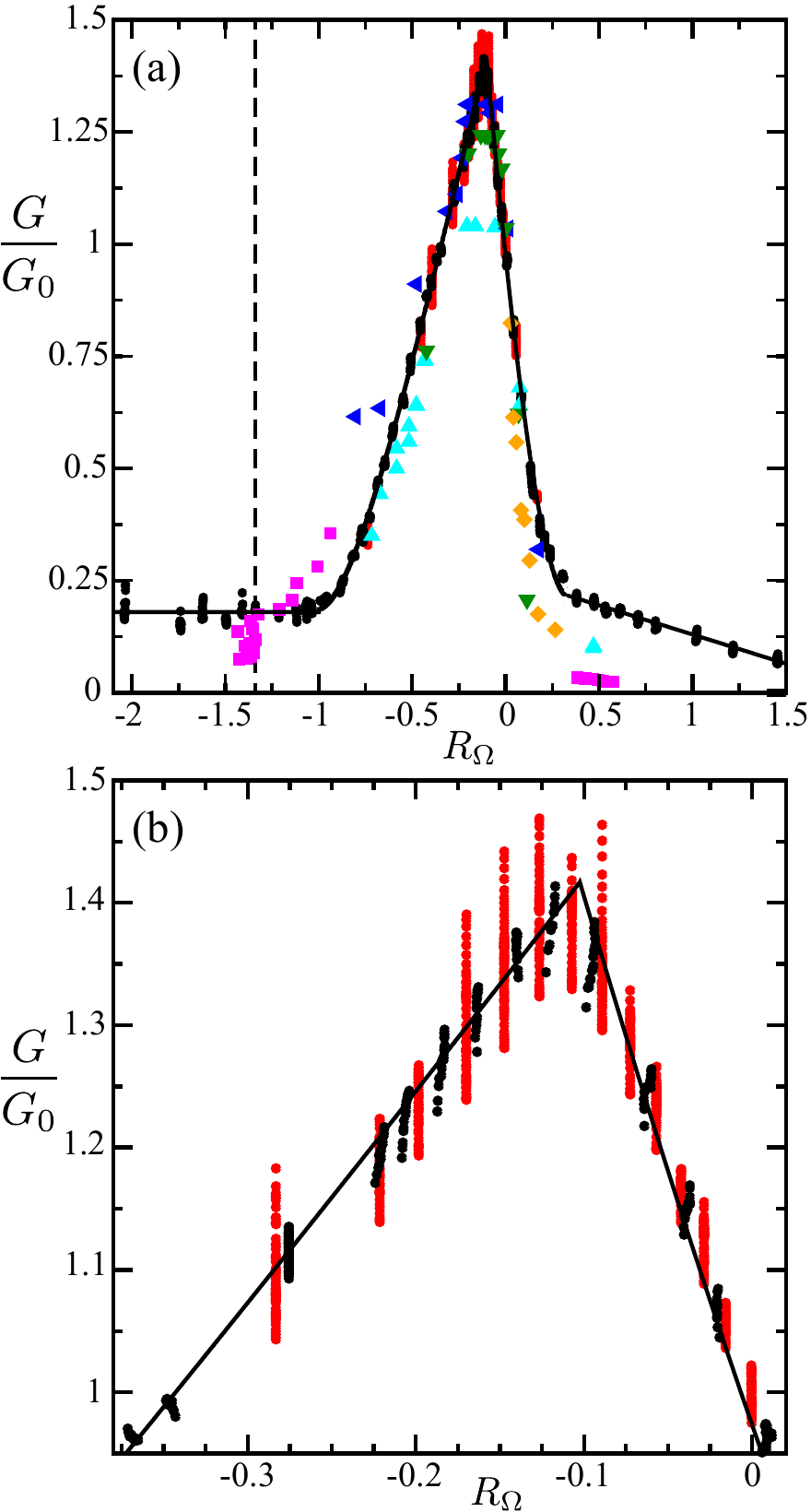}%
\caption{(a) The dimensionless torque $G$ normalized by $G_0 \equiv G(Re, \Rom = 0)$ for different values of the radius ratio $\eta$ nicely collapses
when plotted as function of  $\Rom$.  This normalization is different from Fig.\ 16 in Ref.\ \citep{dubrulle05} where $G$ is normalized by $\Gi \equiv G(Re,\Omega_2 = 0)$.  The data from \citet{wendt33} are shown as cyan ($\eta = 0.68$), blue ($\eta = 0.85$) and green ($\eta = 0.935$) triangles.  Additional radius ratios are examined using the data from \citet{taylor36}, shown as orange diamonds, Richard \citep{richardthesis, dubrulle05}, given by magenta squares, \citet{paoletti11} indicated by black circles and \citet{vangils11} by red circles.  The solid lines are fits given by Eq.\ (\ref{fits}) with smoothened
transitions between the four different regimes.
Keplerian flow states are indicated by the dashed line at $\Rom = -4/3$.
In (b)  we focus on the peak of the curve shown in (a), but only present the data from Twente \citep{vangils11} (red) and Maryland \citep{paoletti11} (black).
The solid line is the fit (\ref{fits}).
}
\label{Gmu_vs_Rom}
\end{center}
\end{figure}

\begin{figure}[tb]
\begin{center}
\includegraphics[width=0.5\columnwidth]{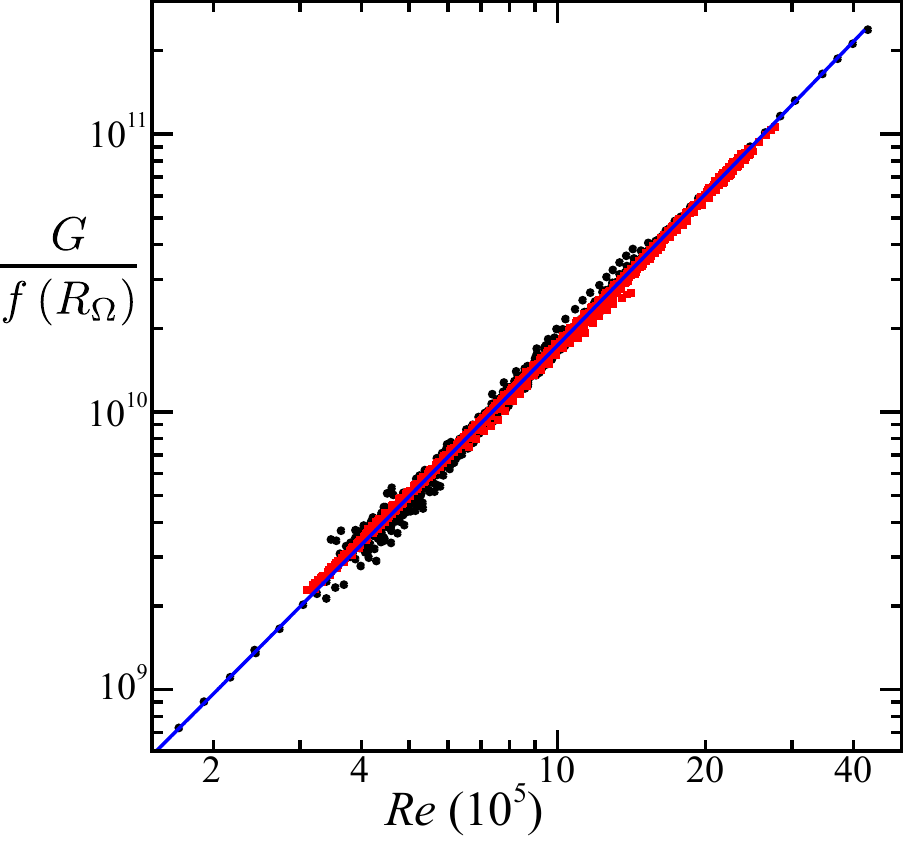}%
\caption{
The dimensionless torque $G$ may be compensated by the fit $f(\Rom)$ given by Eq.\ (\ref{fits}).
The data from both \citet{paoletti11} (black circles) and \citet{vangils11} (red squares) agree with the best fit from \citet{lewis99} for the case where the outer cylinder was stationary (blue line).}
\label{Gcomp_vs_Re_UMD_Twente}
\end{center}
\end{figure}

The agreement between the Twente and Maryland measurements is an important check, but not surprising given that the experiments have very similar geometries, measurement techniques and control parameters (see Table \ref{Table_params}).  Therefore, it remains unclear if the scaling of the normalized torque with $R_\Omega$ is universal or if it depends upon other parameters, such as the radius ratio $\eta$.  While the Twente experiment is capable of examining other values of $\eta$ in future measurements, we may compare our results to previous experiments by \citet{wendt33}, \citet{taylor36} and \citet{richardthesis}, which were analyzed by \citet{dubrulle05}.  The parameters of these past experiments are also summarized in Table \ref{Table_params}.  The measurements of Wendt (triangles), Taylor (orange diamonds) and Richard (magenta squares) are also shown in Fig.\ \ref{Gmu_vs_Rom}(a).

As can be seen in Fig.\ \ref{Gmu_vs_Rom}, the dependence of the normalized torque $G/G_0$ on $\Rom$ for various radius ratios in the range $0.68 \le \eta \le 0.973$ collapse well.   The data appear to follow distinct approximate scalings given by
\begin{equation}
f\(\Rom\) \approx  \left\{
    \begin{array}{ll}
        0.18 \pm 0.06 & \mr{:~}\Rom \le -1\\
        1.79\Rom + 0.05 e^{-4.35\Rom} + 1.59 & \mr{:~}-1 < \Rom \le -0.10\\
        -7.35\Rom + 1.49e^{2.28\Rom} - 0.51 & \mr{:~} -0.10 < \Rom < 0.32\\
        -0.13\Rom + 0.26 & \mr{:}~ \Rom \ge 0.32
    \end{array}
    \right.
    \label{fits}
\end{equation}
Given the rather different span of $Re$ in each experiment and the varying measurement techniques, it is not surprising that a few outliers remain.  For example, the values of $G/G_0$ from Richard's experiments \citep{richardthesis} (magenta squares) were computed using theoretical predictions regarding the correlation between the torque and critical numbers for stability, as described in  \citet{dubrulle05}.  The data from \citet{wendt33}, \citet{taylor36}, \citet{paoletti11} and \citet{vangils11} are direct measurements of the torque, albeit with different apparati and measurement techniques.

\begin{figure}[tb]
\begin{center}
\includegraphics[width=0.5\columnwidth]{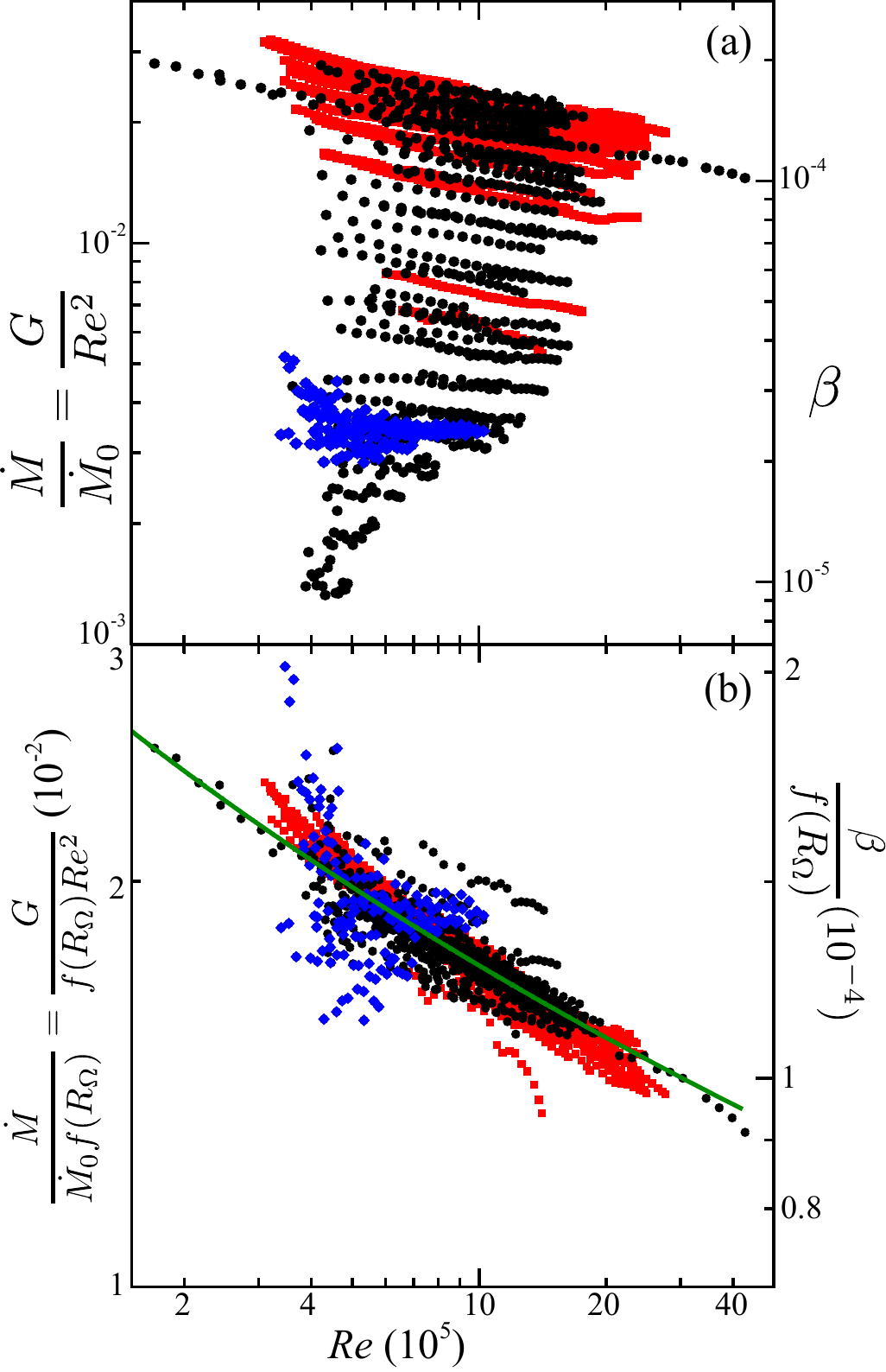}%
\caption{The scaling of (a) $G/Re^2$, which is proportional to the accretion rate $\dot{M}$ and viscosity parameter $\beta$, decays slowly with $Re$ while having an amplitude that varies with the rotation number $\Rom$ (see Fig.\ \ref{Gmu_vs_Rom}).  (b) The effects of global rotation can be accounted for by compensating $G/Re^2$ by the fits $f(\Rom)$ given in Eq.\ (\ref{fits}).  The measurements from Twente are shown as red squares.  The Maryland data are given by black circles, except for quasi-Keplerian flows ($\Rom \le -1$), which are indicated by blue diamonds. The green continuous line is the theoretical fit given in Eq. (\ref{regime3taylor}) normalized by $f(\Rom = -0.2755)$, which corresponds to $\Omega_2 = 0$.}
\label{G_over_Re2_vs_Re}
\end{center}
\end{figure}

\subsubsection{Variation with Reynolds number}

The fits given in Eq.\ (\ref{fits}) may be used to compensate the torque to account for the effects of global rotation.  Figure \ref{Gcomp_vs_Re_UMD_Twente} shows the dimensionless torque $G$ compensated by the fit $f(R_\Omega)$ given in Eq.\ (\ref{fits}) as a function of $Re$.  The data from Twente are shown as red squares while those of Maryland are given by black circles.  In both cases, the data agree well with the best fit given by \citet{lewis99} for the case $\Omega_2 = 0$ (blue line).  This indicates that the amplitude of the torque is affected by global rotation, as measured by $G/G_0$, while the scaling with $Re$ seems to be  independent of $\Rom$ in the regime of our investigation.  Therefore the torque may be approximately factorized and written as
\begin{equation}
G = f(\Rom)g(Re),
\label{G_factor_func}
\end{equation}
where $f(\Rom)$ is measured in Fig.\ \ref{Gmu_vs_Rom} and fit by Eq.\ (\ref{fits}) and our measurements of $g(Re)$, shown in Fig.\ \ref{Gcomp_vs_Re_UMD_Twente}, agree with previous measurements, and may be fit by effective power-laws or logarithmically varying functions such as the Prandtl-\vK form \citep{lathrop92a,lathrop92b,lewis99}.

\subsubsection{Comments}

In addition to the observed data collapse, other interesting features arise in the normalized torque scaling with $\Rom$.  The data from Taylor's experiments \citep{taylor36} were obtained with a stationary inner cylinder $\(\Omega_1 = 0\)$, and the flow is therefore Rayleigh-stable.  However, the data from \citet{wendt33}, Maryland \citep{paoletti11} and Twente \citep{vangils11} over the same range of the rotation number, namely $0 < \Rom < 0.27$, are obtained with counter-rotating cylinders.  This configuration is linearly unstable, yet the measured values of $G/G_0$ agree between all four experiments.  We argue,
as done by  \citet{taylor36}, that the flow states in Taylor's experiments could not have been laminar and must have undergone a subcritical transition to turbulence.

\citet{paoletti11} observed that the torque scaling varied linearly with $\Rom$ in four regions of the parameter space corresponding to: region~I: $\Rom \le -1$, region~II: $-1 < \Rom \le -0.10$, region~III: $-0.10 < \Rom < 0.38$ and region~IV: $0.38 < \Rom$, where we have reordered the region numbers to correspond to increasing $\Rom$.  Figure \ref{Gmu_vs_Rom} shows that all of the data described here follow the region~I--region~II and region~II--region~III crossovers.  The crossover between regions~I and II corresponds to the Rayleigh stability criterion for all values of $\eta$.  The crossover between regions~II and III occurs at the observed maximum in the normalized torque $G/G_0$ as a function of $\Rom$.  While this maximum occurs for all of the \citet{wendt33}, Maryland \citep{paoletti11} and Twente \citep{vangils11} datasets, there is no general theoretical prediction for the exact location of this maximum yet.

The last crossover observed by Paoletti and Lathrop, between regions~III and IV, corresponded to $\Omega_1 = 0$ with $\Omega_1 > 0$ and $\Omega_2 < 0$ in region~III (counter-rotation) to $\Omega_1 > 0$ and $\Omega_2 > 0$ in region~IV (co-rotation).  It is unclear whether
 this crossover is set by $\Rom = 0.38$ or whether it always occurs at $\Rom(\Omega_1 = 0)$.  The data from \citet{taylor36} has $\Omega_1 = 0$ but given the values of $\eta$ used in those experiments $\Rom < 0.38$. The Richard data for $\Rom > 0.38$ fall below the observations by Maryland, although the scaling with $\Rom$ is similar.  It is unclear if this disparity is a result of the analysis used to deduce the torques from Richard's measurements or the difference in the control parameters from the Maryland experiment (see Table \ref{Table_params}).  It would be fruitful to have independent, direct measurements of the torque in region~IV to compare against those of Maryland at different radius ratios to better understand this final crossover.

The Richard and Maryland data both indicate that the observed torque for quasi-Keplerian flows states ($\Rom \le -1$) is approximately 14\% of the maximum observed torque for a given Reynolds number.  This is in contrast to the results of \citet{ji06}, who deduced that such flow states are \lq\lq essentially steady" and unable to transport angular momentum hydrodynamically.  In addition to the different geometries between the experiments in  \citet{ji06} and those of Richard and Maryland (see Table \ref{Table_params}), the measurement techniques and finite-size effects were different between the experiments.  This contrast is further discussed in Section \ref{Discussion}.  Clearly, additional independent measurements of the angular momentum flux for quasi-Keplerian flows could aid this debate, as recently discussed by \citet{balbus11}.

\subsubsection{Astrophysical implications}

As described in Section \ref{Astrophysically relevant quantity}, the scaling of the torque with Reynolds number may be used to determine either the turbulent viscosity parameter $\beta$ or the accretion rate $\dot{M}$ through Eq.\ (\ref{considerons}).  Our measurements of $G/Re^2$, which is proportional to both $\beta$ and $\dot{M}$, are shown in Fig.\ \ref{G_over_Re2_vs_Re}(a).  The measurements from Twente are shown as red squares while those of Maryland are given by black circles.  The data indicated by blue diamonds are quasi-Keplerian flow states ($\Rom \le -1$) from the Maryland experiment.  The quantity $G/Re^2$ decreases slowly with $Re$
(only logarithmically, see \citep{dubrulle05,hersant05,grossmann11}) as expected since $G \propto Re^{\alpha}$ with $1.8 < \alpha(Re) \le 2.0$ for $10^5 < Re < \infty$ \citep{lathrop92a,lathrop92b,lewis99}.  The amplitude of $G/Re^2$ for a given $Re$ is determined by $\Rom$.  The effects of global rotation may be accounted for by compensating the measurements of $G/Re^2$ by $f(\Rom)$ (see. Eq.\ (\ref{fits})), as shown in Fig.\ \ref{G_over_Re2_vs_Re}(b).  As discussed in  \citet{dubrulle02,hersant05}, the scaling may be fit by a logarithmically decaying function, such as:
\begin{equation}
\frac{G}{Re^2}=K_7\frac{\eta^2}{(1-\eta)^{3/2}}\frac{1}{\ln[\eta^2(1-\eta)Re^2/K_8]^{3/2}},
\label{regime3taylor}
\end{equation}
with the values chosen to be $K_7 = 0.4664$ and $K_8 = 10^4$ to describe the best fit for the torque scaling provided by \citet{lewis99} for the case $\Omega_2 = 0$.
An alternative logarithmic dependence between $G/Re^2$ and $Re$
has been suggested by \citet{grossmann11} (for the analogous case of ultimate Rayleigh-B\'enard flow), and it equally well fits the data.

The expected accretion rate, measured by $G/Re^2=3.3\times 10^{-3}$, for quasi-Keplerian flows (blue diamonds in Fig.\ \ref{G_over_Re2_vs_Re}) is approximately 14\% of the maximum accretion rate, which occurs for $\Rom = -0.10$.
This corresponds to a value of $\beta_{Md}=2\times 10^{-5}$ or $\dot M/\dot M_{0}=3.3\times 10^{-3}$ at $Re=7 \times 10^{5}$. Extrapolating towards astrophysical Reynolds numbers by using the formula of \citet{hersant05}, we get a value of $\dot M/\dot M_{0}=10^{-3}$ that is consist with values observed in disks around TTauri stars.

\section{Discussion}
\label{Discussion}

\begin{figure}
\begin{center}
\includegraphics[width=0.5\columnwidth]{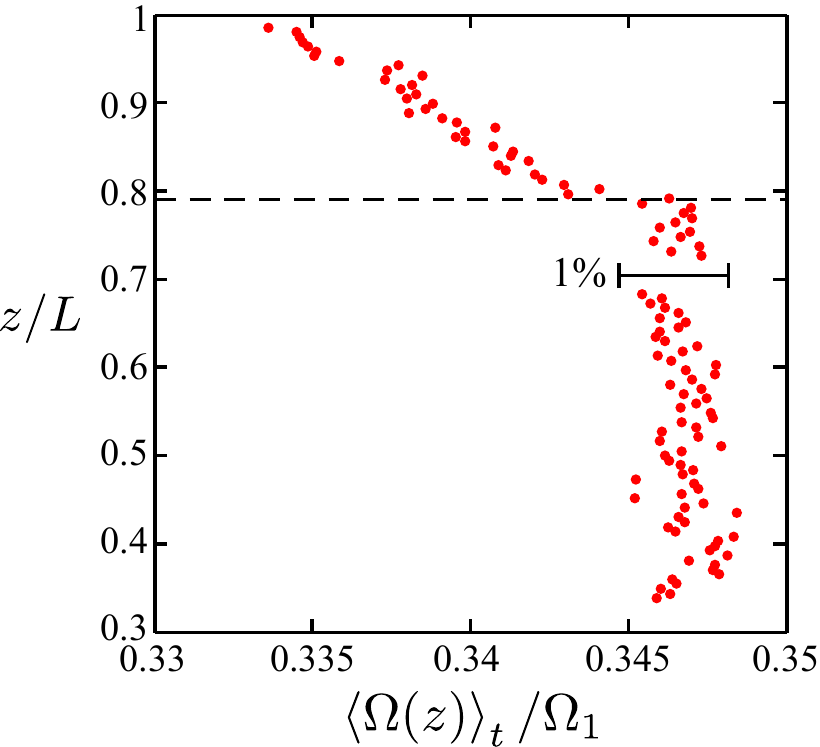}%
\caption{Finite-size effects may be characterized by axial profiles of the angular velocity $\Omega$
 measured at the mid-gap ($r = (a+b)/2$) in the Twente experiment.  In this example, the cylinder Reynolds numbers are $Re_1=1.0\times10^6$ and $Re_2=0$ (outer cylinder stationary). The global torque measurements are taken for $0.22 < z/L < 0.78$ with the top boundary of these measurements indicated by the dashed line at $z/L = 0.78$.  The resolution of the laser-doppler anemometry prohibits measurements of the angular velocity in the boundary layer near $z/L = 1$ where $\Omega =\Omega_2 = 0$, in this case.}
\label{axial-dep}
\end{center}
\end{figure}

The measured accretion rate for all of our quasi-Keplerian flows ($\Rom \le -1$) is approximately 14\% of the maximum accretion rate, which occurs at $\Rom = -0.10$, resulting in $\beta_{\mr{Md}}=2\times 10^{-5}$. This is close to the value computed by Richard and Zahn $\beta_{\mr{RZ}}=10^{-5}$ and to the value of \citet{dubrulle05} $\beta_{\mr{D}}=8\times 10^{-6}$.  This is somewhat surprising given that quasi-Keplerian flow profiles are linearly stable while the maximum accretion rate occurs when the cylinders counter-rotate, which produces destabilizing shear that dominates the dynamics.  This result is in stark contrast to the results presented by \citet{ji06} and \citet{schartman11}, where $\beta_{\mr{P}}$ was measured to be an order of magnitude smaller, specifically $\beta_{\mr{P}}=7 \times 10^{-7}$.  Thus, they concluded that hydrodynamics alone cannot efficiently transport angular momentum in quasi-Keplerian flows.

\citet{schartman11} state that these incompatible results are caused by differing interpretations of finite-size effects, namely Ekman circulation produced by the axial boundaries.  In the Princeton experiments \citep{ji06, schartman09,burin10,schartman11} the aspect ratio is small ($\Gamma \approx 2$) and therefore finite-size effects would likely dominate.  To minimize Ekman circulation, their axial boundaries are split into pairs of rings that can independently rotate with respect to the cylinders.  This method has been shown to reduce the effects of Ekman circulation in the bulk of the fluid \citep{schartman09}.  Furthermore, when Ekman circulation is not minimized their measured values of $\beta$ for quasi-Keplerian flows are approximately half of the typical values presented here.  Thus, \citet{schartman11} argue that our larger values of $\beta$ are the result of Ekman circulation that is present in other experiments, but is minimized in their apparatus.

We claim, on the other hand, that our measured torques cannot be solely attributed to finite-size effects.  The main premise of our argument is that the normalized torques $G/G_0$ agree quite well for all of the different experimental data detailed here, even though the apparatus geometry, measurement techniques and control parameters vary greatly.  Given this widespread disparity in geometry, rotation rates and range of Reynolds numbers, one would expect that the effects of Ekman circulation would differ and result in systematic discrepancies between the experiments.  We do not observe such discrepancies and we further detail our arguments along these lines below.

The torques in the Maryland and Twente experiments are measured over only 56\% of the axial length of the flow centered at the mid-height of the apparatus.  This design has been intentionally implemented such that the torque measurements are unaffected by flows within 2.58 radial gap widths of either axial boundary where Ekman circulation is strongest.  Wendt's experiments \citep{wendt33} had a free upper surface, which is therefore devoid of Ekman circulation.  Even though Taylor's measurements \citep{taylor36} were affected by Ekman circulation for $\Gamma > 100$, his measurements of the torque for Rayleigh-stable flows that are expected to be most affected by Ekman circulation agree with our measurements of counter-rotating flows for the same value of $\Rom$, where Ekman circulation is dwarfed by the dominant shear.

One of course wonders on whether this Ekman circulation caused  by the co-rotating upper and lower plates affect the flow patterns.
For laminar flow e.g. for purely outer cylinder rotation this clearly is the case \citep{coles66,hollerbach04}. However, we argue that the strong turbulence destroys such an axial dependence of the flow field. To check this assumption, we have measured the full turbulent velocity profile with Laser Doppler Anemometry (LDA) \citep{vangils12}.  The result is shown in Fig.\ \ref{axial-dep}.  Evidently, over the length of the central section of the inner cylinder where we measure the torque there is no visible $z$-dependence of the mean velocity; the plate boundary layers only produce an axial dependence far above the domain of the central section of the inner cylinder.

\citet{ji06} stated that their measured velocity fluctuations and values of $\beta$ were much higher for more viscous fluids, even though the flow profiles were the same.  They interpreted this observation to imply that the residual finite-size effects penetrate deeper into the bulk of the fluid at lower $Re$, even though Ekman circulation was claimed to have been minimized in both sets of measurements thereby resulting in similar (nearly ideal-Couette) flow profiles.  The values of $Re$ for the experiments presented here span two orders of magnitude while the values in viscosity vary by a factor of 34, yet no such systematic effects on the measured torques have been observed.  Furthermore, given the small aspect ratio of the Princeton experiments one would expect Ekman circulation to produce significantly higher values of $\beta$ when the axial boundaries co-rotate with the outer cylinder, as in the experiments presented here.  However, even in this case their measured values are a factor of two smaller than those presented here and in \citet{paoletti11}.

The normalized torque $G/G_0$ is approximately constant for each value of $\Rom$ characterized by the present experiments.  This indicates that the amplitude of the torque varies with $\Rom$, however the scaling with $Re$ is unchanged (see Fig.\ \ref{Gcomp_vs_Re_UMD_Twente}).  \citet{paoletti11} observed that their torque scaling for Rayleigh-stable and unstable flows agreed with the \PvK description of \citet{lathrop92b,lewis99}.  The Twente data, initially presented by  \citet{vangils11}, was well-described by the theoretical predictions of \citet{eckhardt07a} for all rotation numbers explored.  If Ekman circulation significantly contributed to the torque for Rayleigh-stable flows, as argued by \citet{schartman11}, then one would expect systematic disparities between these theoretical descriptions and the measured torques.  As shown in Fig.\ \ref{Gcomp_vs_Re_UMD_Twente}, though, the torque scaling agrees well with the observations of \citet{lewis99} for the case $\Omega_2 = 0$ \textit{for all values of $\Rom$}.

The Princeton experiments \citep{ji06,schartman09,burin10,schartman11} measure \textit{local velocity correlations to determine the global angular momentum flux}, whereas the Maryland and Twente experiments directly measure the global angular momentum flux using precision calibrated force-arm measurements.  The distinction between the measured quantities has made direct comparisons of the experiments difficult.  To bridge this gap, \citet{burin10} recently used their local velocity measurements to determine the corresponding torque $G$.  To compare to previous studies with $\Omega_2 = 0$ \citep{lathrop92a,lathrop92b,lewis99} \citet{burin10} fit the scaling of $c_f^{-1/2}$ to $\log_{10}(Re*c_f)$, where $c_f \equiv G/Re^2$, and determined a slope of $1.40 \pm 0.16$.  The authors claim that this value agrees with the measured value of 1.52 reported by  \citet{lathrop92a,lathrop92b} and 1.56 measured in  \citet{lewis99}.  However, the appropriate \PvK scaling predicts the following relationship \citep{lathrop92a,lathrop92b,lewis99}
\begin{equation}
\frac{1}{\sqrt{c_f}} = N \log_{10}(Re\sqrt{c_f}) + M,
\end{equation}
in contrast to the scaling shown in Fig.~5 of  \citet{burin10}.  Furthermore, the fit parameters $N$ and $M$ are predicted to depend upon the radius ratio in the following way
\begin{eqnarray}
N = [(1-\eta^2)\ln 10]/(\eta\kappa\sqrt{2\pi}), \\
M = \frac{N}{\ln 10}\[\ln \left\{\(\frac{1-\eta}{1+\eta}\)\frac{1}{y_0^+\sqrt{2\pi}}\right\} + \kappa y_0^+\],
\end{eqnarray}
where $\kappa = 0.40$ is the \vK constant and $y_0^+ = 5$.  The radius ratio $\eta = 0.35$ in the Princeton experiments, therefore the \PvK skin friction law predicts $N = 5.76$ and $M = -3.15$.  These values do not agree with our analysis of the data in \citet{burin10} where $N \approx 2$ and $M \approx -0.4$.  In fact, the measured values of $G$ determined by the local velocity fluctuations are on average five times larger than those predicted by the \PvK skin friction law that has been verified using direct torque measurements \citep{lathrop92a,lathrop92b,lewis99,paoletti11}.  Therefore, we argue that the local velocity fluctuation measurements used in the Princeton experiments \citep{ji06,schartman09,burin10,schartman11} face severe experimental challenges.
In fact, \citet{huis11}  showed  that the local angular momentum transport undergoes fluctuations that are two orders of magnitude larger than the
mean value. However, once the local angular momentum transport is averaged long enough and over a large enough region, \citet{huis11} achieved
perfect agreement between the local and the global measurements.

Clearly the contrasting conclusions of the Princeton experiments \citep{ji06,schartman09,burin10,schartman11} and those presented here warrant further experimental and theoretical investigation, as suggested by \citet{balbus11}.  It may be  that the Princeton measurements of the local velocity fluctuations
 may not accurately determine the global angular momentum transport (\citep{burin10}), due to lack of convergence.  On the other hand, the effects of the axial boundaries on the global torque measurements presented here have not yet been characterized in detail.  The Twente experiments \citep{huis11}
provide both  local velocities as well as the global torque, which thereby presents  future opportunities to clarify these distinctions.
Vice versa,
direct global torque measurements in a split axial-ring apparatus, such as the Princeton device, would also be scientifically useful.

\section{Conclusions}
\label{Conclusions}

The measurements that we present here indicate that the dimensionless torque $G$ may be described as a product of functions that separately depend upon the Reynolds number $Re$ and the rotation number $\Rom$.
The rotation number $\Rom$ introduced by \citet{dubrulle05} collapses all of the experimental data presented here for various radius ratios, aspect ratios and Reynolds numbers (see Table \ref{Table_params}).  The effects of global rotation, measured by $G/G_0$, where $G_0 \equiv G(Re,\Rom = 0)$, are shown to be well-described by the function $f(\Rom)$ given in Eq.\ (\ref{fits}) for the entire range of $\Rom$ explored in the present studies.

The most astrophysically relevant result is the scaling of $G/Re^2$, which is proportional to both the turbulent viscosity parameter $\beta$ and the accretion rate $\dot{M}$.  We observe that the expected accretion rate for quasi-Keplerian flows ($\Rom \le -1$) is approximately 14\% of the maximum attainable rate for a given Reynolds number.  We therefore get a dimensionless angular transport parameter $\beta_{\mr{Md}}=2\times 10^{-5}$, in agreement with Richard and Zahn but more than an order of magnitude larger than the Princeton experiments. Such angular momentum transport is able to produce dimensionless accretion rates on the order of $\dot{M}/\dot{M_0} \sim 10^{-3}$, which are compatible with observations in disks around TTauri stars \citep{hersant05}.  We have argued that our results are not easily attributed to finite-size effects, such as Ekman circulation.  As such, we conclude that hydrodynamics may be able to transport angular momentum at astrophysical rates in spite of the linear stability provided by the radially-increasing angular momentum profile.  This level of transport must result from turbulent means, thereby implying that quasi-Keplerian flows can be nonlinearly unstable to finite amplitude disturbances.

\begin{acknowledgements}
The Maryland group would like to thank B. Eckhardt, Michael E. Fisher, C. Kalelkar, D. Martin, Harry L. Swinney and D. S. Zimmerman and acknowledge the support of Grant No. NSF-DMR 0906109.  BD thanks the CNRS for its continuous unconditional financial support that made this fundamental research possible.  The Twente collaboration thanks G. W. Bruggert for crucial work in building the $T^3C$ device,
S.\ Grossmann and S. G. Huisman for discussions, and STW, which is financially supported by NWO, for financial
support of this project.
\end{acknowledgements}

\bibliographystyle{aa} 

\end{document}